\documentclass[conference,table]{IEEEtran}
\IEEEoverridecommandlockouts
\usepackage{bbding}
\usepackage{cite}
\usepackage{amsmath,amssymb,amsfonts}
\usepackage{bm} 
\usepackage{textcomp}
\usepackage{caption}
\usepackage{graphicx}
\usepackage{pgfplots}
\pgfplotsset{compat=1.18}
\usepackage{subcaption}
\usepackage{multirow}
\usepackage{algorithmic}
\usepackage{algorithm}
\pgfplotsset{ 
cycle list={%
{draw=black,mark=star,solid},
{draw=black, mark=square,solid}}}
\usepackage{breqn}
\usepackage{xcolor}
\usepackage{tikz}
\usetikzlibrary{plotmarks}
\definecolor{darkblue}{rgb}{0.0, 0.0, 0.55}
\definecolor{magenta}{rgb}{0.79, 0.08, 0.48}
\usepackage{caption}
\usepackage{comment}
\usepackage{authblk}
\usepackage{fancyhdr}
\usepackage{subcaption}
\usepackage{tabularx} 
\usepackage{hyperref}
\usepackage{booktabs}
\usepackage[table]{xcolor}
\usepackage{makecell}
\usepackage{fancyhdr}
\fancypagestyle{firstpage}
{
    \fancyhead[L]{\footnotesize © 2026 IEEE.  Personal use of this material is permitted.  Permission from IEEE must be obtained for all other uses, in any current or future media, including reprinting/republishing this material for advertising or promotional purposes, creating new collective works, for resale or redistribution to servers or lists, or reuse of any copyrighted component of this work in other works. This paper is accepted at IEEE International Symposium on Circuits and Systems (ISCAS) 2026.}
    \fancyhead[R]{}
}

\begin{document}

\title{ An FPGA-Based SoC Architecture with a RISC-V Controller for Energy-Efficient Temporal-Coding Spiking Neural Networks\\
}

\author[1]{Mohammad Javad Sekonji}
\author[1]{Ali Mahani}
\author[1]{Maryam Mirsadeghi}
\author[2,3]{Mahdi Taheri}

\affil[1]{Shahid Bahonar University of Kerman, Kerman, Iran}
\affil[2]{Brandenburg Technical University, Cottbus, Germany}
\affil[3]{Tallinn University of Technology, Tallinn, Estonia}

\maketitle
\thispagestyle{firstpage}

\begin{abstract}
Spiking Neural Networks (SNNs) offer high energy efficiency and event-driven computation, ideal for low-power edge AI. Their hardware implementation on FPGAs, however, faces challenges due to heavy computation, large memory use, and limited flexibility. This paper proposes a compact System-on-Chip (SoC) architecture for temporal-coding SNNs, integrating a RISC-V controller with an event-driven SNN core. It replaces multipliers with bitwise operations using binarized weights, includes a spike-time sorter for active spikes, and skips non-informative events to reduce computation. The architecture runs fully on a Xilinx Artix-7 FPGA, achieving up to 16× memory reduction for weights and lowering computational overhead and latency, with 97.0\% accuracy on MNIST and 88.3\% on Fashion-MNIST. This self-contained design provides an efficient, scalable platform for real-time neuromorphic inference at the edge.
\end{abstract}


\section{Introduction}
Neuromorphic computing has emerged as a promising paradigm for brain-inspired processing, offering energy-efficient and event-driven computation. Among its models, Spiking Neural Networks (SNNs) stand out for encoding and processing information via discrete spike events, mimicking biological neurons. Temporal-coding SNNs, in particular, leverage spike timing for sparse activation, reducing computational costs compared to conventional Artificial Neural Networks (ANNs) \cite{schuman2017, zhao2024, sparq}.

Despite their potential, efficient hardware implementation of SNNs remains challenging. High-precision weights and complex arithmetic in synaptic operations drive energy consumption and silicon area, posing issues for resource-constrained edge devices~\cite{B1}. While neuromorphic accelerators like IBM's TrueNorth and Intel's Loihi \cite{2014truenorth, 2018loihi} offer efficient spike processing, their fixed architectures limit flexibility. FPGAs, however, provide reconfigurability and cost-effectiveness, making them suitable for prototyping diverse SNN models on edge platforms. For instance, the SYNtzulu processor~\cite{syntzulu2025} demonstrates an event-driven FPGA design. However, its optimization for continuous dynamic signals (e.g., sEMG) using delta modulation makes it unsuitable for the frame-based, temporal-coded inference needed for image classification.

The demand for lightweight SNNs has spurred interest in Binarized Neural Networks (BNNs), which replace high-precision weights with binary values (±1) to minimize memory and computation needs—critical for edge hardware \cite{zhu2023spikegpt}. The BS4NN model exemplifies this, integrating binarized weights with temporal coding to match full-precision SNN accuracy while reducing resource demands \cite{BS4NN2022}. Yet, BS4NN evaluations have been software-based, lacking a hardware architecture to fully exploit its efficiency.

\begin{figure*}[htbp]
\centerline{\includegraphics[width=\textwidth]{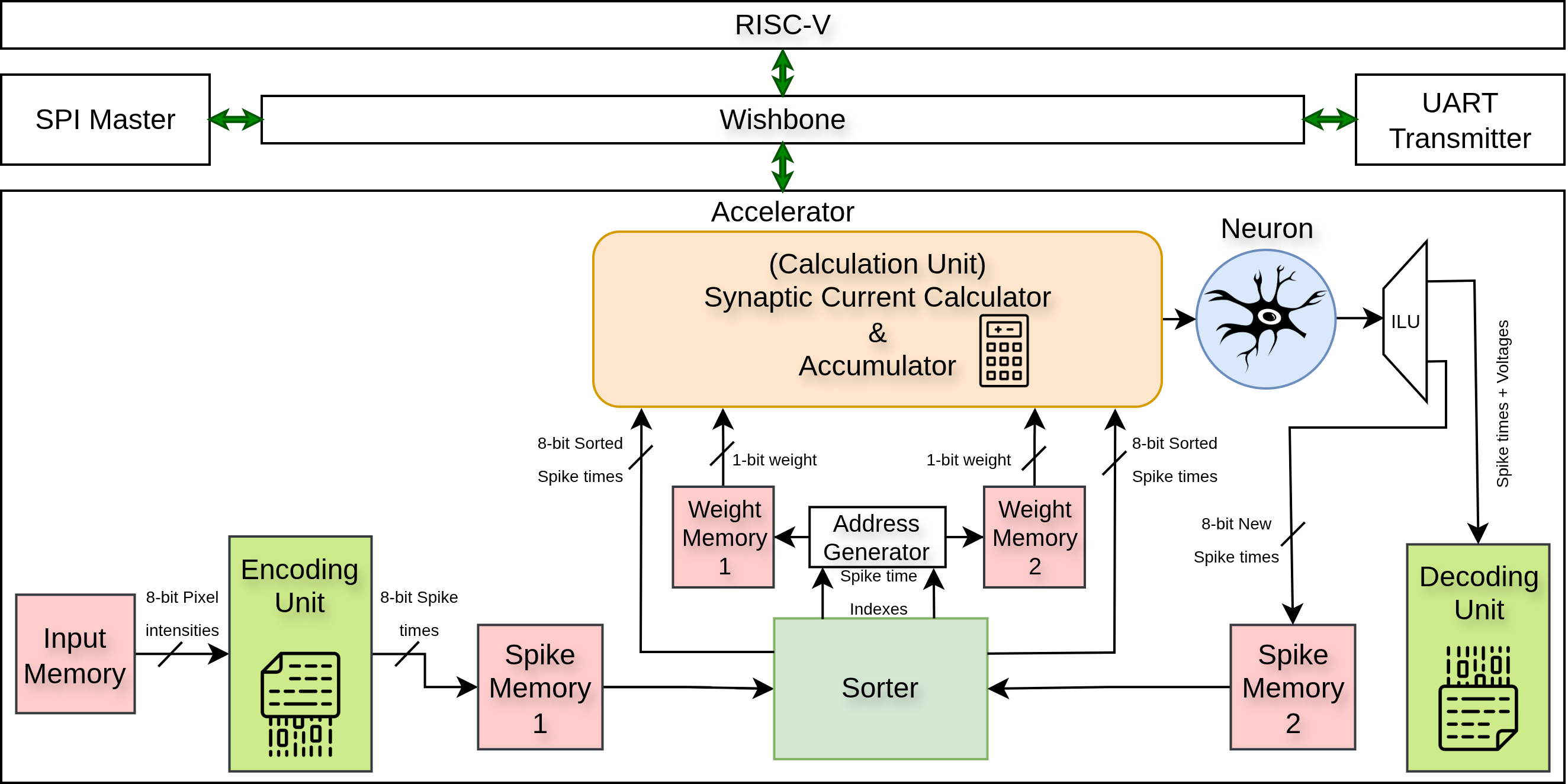}}
\caption{Proposed SNN hardware architecture, integrating a RISC-V controller (upper section) and a dedicated SNN processing core (lower section) with labeled components.}
\label{HW_Architecture}
\end{figure*}

Existing FPGA-based SNN implementations, especially for BSNNs, face limitations: reliance on external processors, sequential spike processing, and inefficient on-chip data handling hinder real-time, autonomous inference. To address this, we propose an energy-efficient FPGA-based System-on-Chip (SoC) for temporal-coding SNNs, integrating a RISC-V controller with an SNN core (Fig.~\ref{HW_Architecture}). The controller manages configuration and memory, while the core executes spike computation via modular components (encoder, sorter, synaptic calculator, neuron array, decoder). This design supports both binary and fixed-point weighted SNNs, adapting to various configurations without hardware redesign. It achieves up to 16× memory reduction with binarized weights, processes only active spikes, skips non-informative events for efficiency, and offers low power consumption with minimal resource usage.

The remainder of this paper is organized as follows. Section II details the proposed SoC architecture and its optimizations. Section III presents the experimental setup and performance evaluation on benchmark datasets. Finally, Section IV concludes the paper, summarizing contributions and outlining future work.

 \begin{figure*}[htbp]
 \centerline{\includegraphics[width=\textwidth]{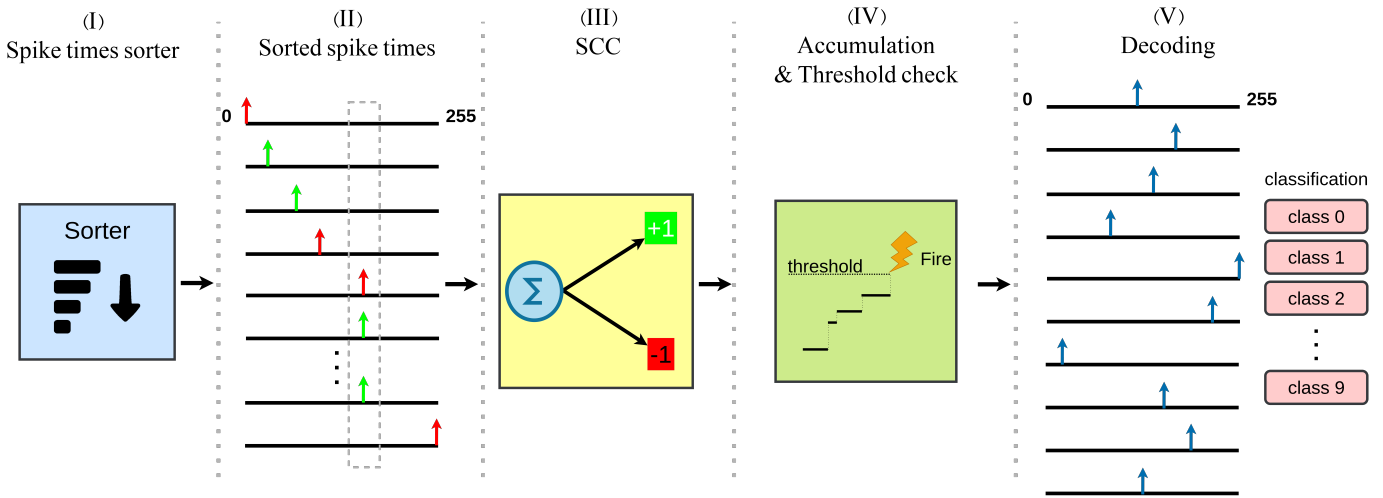}}
\caption{Proposed SNN computational flow: (I) Spike sorting and event selection; (II) Binary-weight accumulation; (III) Synaptic current calculation; (IV) Neuron integration and firing; (V) Output decoding and classification.}
\label{snn_process}
\end{figure*}

\section{Proposed Hardware Architecture}

The proposed hardware architecture is developed as an energy-efficient FPGA-based System-on-Chip (SoC) designed to accelerate temporal-coding Spiking Neural Networks (SNNs) with minimal hardware resources. As illustrated in Fig.~\ref{HW_Architecture}, the architecture integrates two primary components: a RISC-V controller and a dedicated SNN processing core. The RISC-V processor orchestrates configuration, data flow, and memory management, while the SNN core, detailed in the figure’s lower section, executes spike-based inference through specialized modules. This co-design provides a self-contained, programmable, and highly efficient SNN platform suitable for low-power devices.

\subsection{System Overview and RISC-V Controller}
The RISC-V controller configures the SNN parameters and manages on-chip BRAMs. It supports the standard RISC-V GNU compiler toolchain~\cite{riscv-gnu-toolchain}, enabling rapid reconfiguration without hardware re-synthesis. 
An \texttt{RV32I}-based controller fetches input data and pre-trained weights from external SPI Flash, stores them on-chip, and triggers SNN inference via interrupts. After label determination, a completion interrupt initiates loading of the next input sample.
Inference results are transmitted via UART, forming a standalone neuromorphic platform.

\begin{figure}[t]
	\captionsetup{labelfont=,textfont=}
	\centering
	\includegraphics[width=\columnwidth]{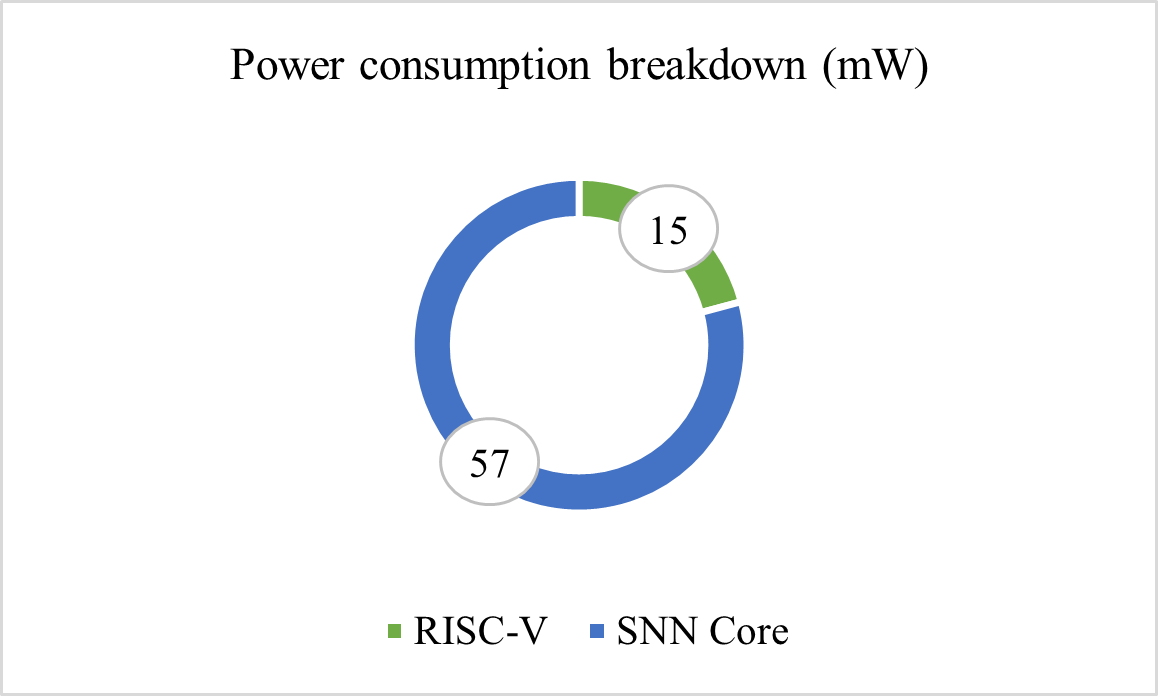}
	\caption{Power consumption breakdown of architecture.}
	\label{fig:power}
\end{figure}

\begin{figure}[t]
	\captionsetup{labelfont=,textfont=}
	\centering
	\includegraphics[width=\columnwidth]{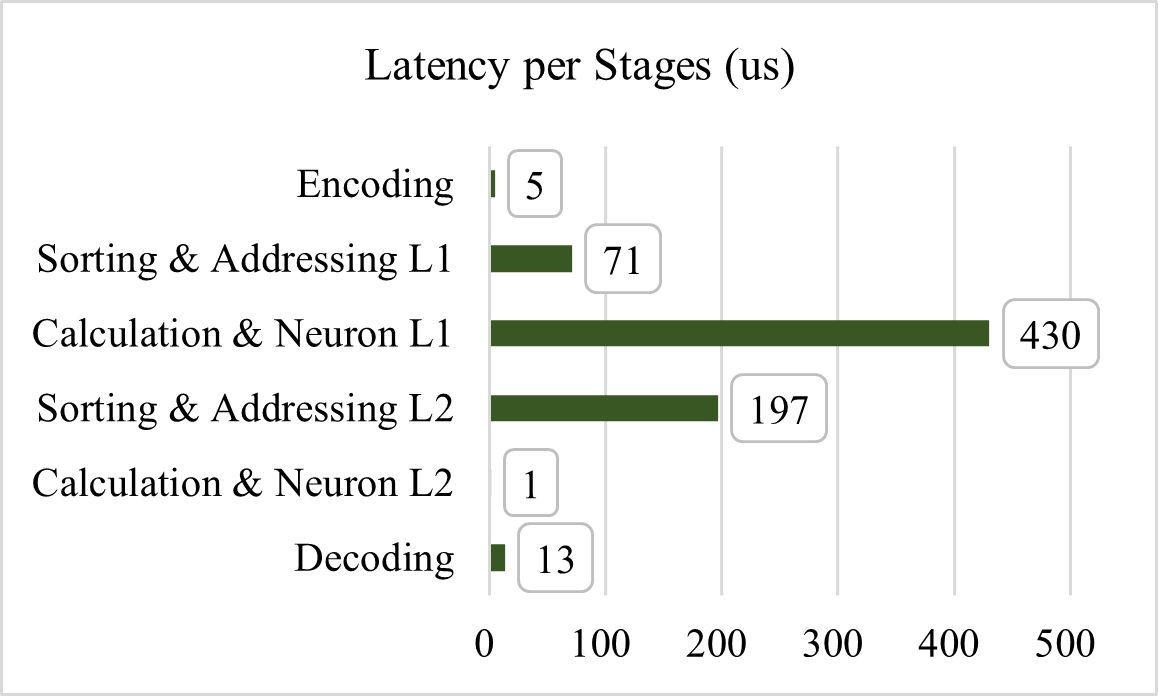}
	\caption{Latency breakdown of pipeline stages.}
	\label{fig:latency}
\end{figure}

\subsection{Encoding and Spike Generation}
After the input samples are stored in memory by the controller, they are sequentially fed into the Encoding Unit, which consists of an input buffer and a Time-to-First-Spike (TTFS) encoder. The TTFS encoder transforms raw pixel intensities into temporal spike patterns, where higher input values produce earlier spikes within a fixed time window.
From the hardware perspective, this mapping is efficiently implemented through a simple inverted logic (NOT) operation, realizing the intensity-to-delay transformation with negligible latency and hardware cost.


\subsection{Data Flow and Pipeline Processing}
The SNN core processes encoded spikes through a layer-wise pipeline, including spike memories, weight memories, sorting units, synaptic current calculators (SCC), neurons, and decoding units, as depicted in Fig.~\ref{snn_process}. 
Spike memories store 8-bit temporal codes, while weight memories hold binarized weights ($\pm 1$), packing 16 weights per 16-bit cell. 
The spike sorter arranges active events by timing and passes them to the synaptic current calculator (SCC),
featuring a modular and scalable structure that accommodates increasing input activity, while potentially becoming timing-critical at higher input densities.
The synaptic current calculator computes currents via lightweight addition/subtraction:
\begin{equation}
	I(t) = \alpha^l \sum_i B_{ji}^l S_i^{l-1}(t),
	\label{eq:current_short}
\end{equation}
where $B_{ji}^l \in \{-1,+1\}$. Multipliers are thus replaced by bitwise operations, eliminating DSP usage. The accumulated current is processed by a configurable non-leaky Integrate-and-Fire neuron, which emits a spike once its membrane voltage surpasses a threshold. Spike times are then propagated to the next layer. In the final layer, a decoding unit determines the predicted class based on the earliest spike or, if silent, the maximum membrane potential. The output is transmitted via UART, completing the inference process.

\begin{table*}[t]
	\centering
	\caption{Performance Comparison of Proposed SNN Implementations with a Prior Artix-7 Design}
	\label{tab:snn_comparison}
	\footnotesize
	\begin{tabular}{lccccccccccc}
		\toprule
		\textbf{Hardware} & \textbf{Arch.} & \textbf{Acc. (\%)} & \textbf{Freq. (MHz)} & \textbf{Power (mW)} & \textbf{Latency (ms)} & \textbf{Throughput (fps)} & \textbf{DSP} & \textbf{BRAM} & \textbf{FF} & \textbf{LUT} \\
		\midrule
		Artix-7 (2022) \cite{47} & 400-10 & 73.96 & 100 & 60465 & 0.215 & 4651 & 64 & 45 & 26853 & 29145 \\
		Artix-7 (2025) \cite{51} & 128-10 & - & 100 & 381 & 0.52 & 1923 & - & 44.5 & - & 13464 \\
		\rowcolor{gray!10}
		\makecell[l]{\textbf{Artix-7 (Proposed)} \\ \scriptsize \textbf{[On MNIST]}}  & \textbf{600-10} & \textbf{97.0} & \textbf{163} & \textbf{72} & \textbf{0.718} & \textbf{1392} & \textbf{1} & \textbf{19.5} & \textbf{15591} & \textbf{14164} \\
		\rowcolor{gray!10}
		\makecell[l]{\textbf{Artix-7 (Proposed)} \\ \scriptsize \textbf{[On Fashion-MNIST]}}  & \textbf{1000-10} & \textbf{88.3} & \textbf{167} & \textbf{158} & \textbf{2.64} & \textbf{378}  & \textbf{2} & \textbf{35.5} & \textbf{19485} & \textbf{18404} \\
		\bottomrule
	\end{tabular}
\end{table*}

\begin{table*}[t]
	\centering
	\caption{Performance Metrics for S4NN on the Proposed Hardware}
	\label{tab:nonbinary}
	\footnotesize
	\begin{tabular}{lccccccccccc}
		\toprule
		\textbf{Hardware} & \textbf{Arch.} & \textbf{Accuracy (\%)} & \textbf{Frequency (MHz)} & \textbf{Power (mW)} & \textbf{DSP} & \textbf{BRAM} & \textbf{FF} & \textbf{LUT} & \textbf{Throughput (fps)} \\
		\midrule
		S4NN & 400-10 & 97.4 & 163 & 121 & 1 & 165 & 15654 & 14445 & 2427 \\
		\bottomrule
	\end{tabular}
\end{table*}

\begin{table*}[t]
	\centering
	\caption{Comparison of Binary and Non-Binary Hardware - BS4NN and S4NN}
	\label{tab:binary_vs_nonbinary}
	\footnotesize
	\begin{tabular}{lccccccccccc}
		\toprule
		\textbf{Hardware} & \textbf{Arch.} & \textbf{Weight Precision} & \textbf{Frequency (MHz)} & \textbf{Power (mW)} & \textbf{DSP} & \textbf{BRAM} & \textbf{FF} & \textbf{LUT} & \textbf{Throughput (fps)} \\
		\midrule
		BS4NN & 128-10 & 1 & 163 & 70 & 0 & 7.5 & 15510 & 12812 & 5103 \\
		S4NN & 128-10 & 16 & 163 & 138 & 1 & 68 & 15542 & 13389 & 5696 \\
		\bottomrule
	\end{tabular}
\end{table*}

\subsection{Flexibility and Adaptability to Non-Binary Models}
\label{flexibility}
The proposed SoC architecture is designed to support not only binarized SNNs but also non-binary models by incorporating a flexible weight representation module. This module allows switching between binary (±1) and multi-bit fixed-point weights, enabling compatibility with models like S4NN~\cite{s4nn2020}. Adjustments include reconfiguring the synaptic current calculator (SCC) to handle multi-bit arithmetic and expanding weight memory to accommodate higher precision, ensuring minimal redesign effort while maintaining performance across diverse network configurations.

\subsection{Optimization and Efficiency}
The architecture enhances efficiency and scalability by using binary weight representation to reduce memory usage and employing event-driven processing with a sorter to prioritize active spikes. Skipping non-informative spikes lowers latency without accuracy loss, while bitwise operations minimize DSP usage. Implemented on a Xilinx Artix-7 FPGA, the SoC delivers efficient inference and robust generalization across datasets.

\section{Experimental Setup and Evaluation on Hardware}
To demonstrate the efficacy of the proposed FPGA-based SoC, it was evaluated using two prominent temporal-coding SNN benchmarks: the binarized BS4NN~\cite{BS4NN2022} for performance validation, and the real-valued S4NN~\cite{s4nn2020} to showcase flexibility and modularity. Both models, widely regarded as foundational for single-spike temporal learning, were trained offline. Evaluations were conducted on the MNIST and Fashion-MNIST datasets using the Xilinx Artix-7 FPGA, with two distinct hardware architectures to demonstrate the scalability and generality of the proposed architecture, with parameters aligned with their software counterparts. The trained weights were loaded via the RISC-V controller for fully autonomous on-chip inference, enabling direct comparison with software results.


\subsection{Performance Metrics and Analysis}
The proposed FPGA-based SoC demonstrates excellent performance across two configurations: a 600-10 network achieving 97.0\% accuracy on the MNIST dataset, and a 1000-10 network reaching 88.3\% accuracy on the Fashion-MNIST dataset, both matching software-level results. The architecture delivers low power consumption and optimized resource usage, as detailed in Table~\ref{tab:snn_comparison}. This efficiency is driven by the time-multiplexed neuron module, which reduces logic utilization by nearly 50\% compared to parallel designs~\cite{47}, minimizing inter-neuron communication overhead while maintaining high accuracy. 
Fig.~\ref{fig:power} further analyzes the power consumption distribution, showing that the SNN processing core dominates the overall power budget, while the \texttt{RV32I}-based RISC-V controller contributes only a minor fraction, confirming the effectiveness of the lightweight control-oriented design.
These results highlight the architecture's effectiveness for energy-efficient, real-time SNN inference on resource-constrained edge devices, outperforming~\cite{47} in power and accuracy, and surpassing~\cite{51} with significantly lower power (72 mW vs. 381 mW) and comparable latency.
Additionally, the latency breakdown across pipeline stages is illustrated in Fig.~\ref{fig:latency}, where the cumulative delay of encoding, sorting, neuron computation, and decoding stages corresponds to the total inference latency reported in Table~\ref{tab:snn_comparison}.

\subsection{Flexibility and Non-Binary Model Support}
The proposed architecture’s flexibility, detailed in Section~\ref{flexibility}, enables seamless support for non-binary SNN models like S4NN~\cite{s4nn2020} without hardware redesign. As shown in Table~\ref{tab:nonbinary}, the S4NN implementation on this platform achieves 97.4\% accuracy on the MNIST dataset, closely matching its software baseline at 163 MHz. This performance, with 121 mW power consumption and efficient use of minimal resources, highlights the architecture’s ability to handle multi-bit fixed-point weights effectively, enhancing its versatility for diverse SNN applications.

\subsection{Hardware Optimization: BS4NN vs. S4NN}
For a fair hardware comparison, BS4NN and S4NN were implemented with an identical 128-10 architecture, as shown in Table~\ref{tab:binary_vs_nonbinary}. The BS4NN’s use of binarized weights (±1) reduces memory usage by up to 16× compared to S4NN’s 16-bit weights, lowering BRAM consumption (7.5 vs. 68) and eliminating DSP usage (0 vs. 1), while cutting power consumption (70 mW vs. 138 mW). 
When normalized to comparable network configurations, both designs also demonstrate higher throughput than prior hardware implementations in this domain, highlighting the efficiency of the proposed architecture beyond resource and power savings.
This demonstrates the efficiency of binarization for resource-constrained edge devices, enhancing neuromorphic hardware design.

\section{Conclusion}
This work advances neuromorphic computing with an energy-efficient FPGA-based System-on-Chip (SoC) tailored for temporal-coding Spiking Neural Networks (SNNs). Integrating a RISC-V controller with a dedicated SNN core, the design reduces memory usage by up to 16$\times$ through \textbf{binary weights} and event-driven processing, eliminating multipliers and enabling low-power operation (72 mW) on FPGA. Its flexibility supports both binarized and non-binary SNNs, adapting to BS4NN and S4NN frameworks without hardware redesign. A \textbf{time-multiplexed neuron module} cuts logic utilization by nearly 50\%, enhancing efficiency. Experimental results confirm real-time inference, making it ideal for edge devices. 

Future work will focus on enabling on-chip learning capabilities to support training directly on the proposed architecture, eliminating the need for off-chip training and enhancing system autonomy. In addition, the architecture will be extended to efficiently support deeper and more complex SNN topologies, paving the way for scalable deployment in more demanding neuromorphic applications.

\section*{Acknowledgements}
\scriptsize
This work was supported in part by the Estonian Research Council grant PUT PRG1467 "CRASHLESS“, EU Grant Project 101160182 “TAICHIP“, by the Deutsche Forschungsgemeinschaft (DFG, German Research Foundation) – Project-ID "458578717", and by the Federal Ministry of Research, Technology and Space of Germany (BMFTR) for supporting Edge-Cloud AI for DIstributed Sensing and COmputing (AI-DISCO) project (Project-ID "16ME1127").

\bibliographystyle{ieeetr} 
\bibliography{ref}  

\end{document}